\documentclass[12pt]{article}
\usepackage{epsfig}
\usepackage{a4,isolatin1}
\usepackage{amsmath,amsfonts,latexsym, amssymb}
\newtheorem{satz}{Theorem}[section]
\newtheorem{defi}[satz]{Definition}

\newtheorem{bem}[satz]{Remark}

\newtheorem{koro}[satz]{Corollary}
\newtheorem{bsp}[satz]{Example}
\newtheorem{assumption}[satz]{Assumption}
\newtheorem{obdef}[satz]{Observation/Definition}
\newtheorem{conclusion}[satz]{Conclusion}
\newtheorem{ob}[satz]{Observation}
\newtheorem{consequences}[satz]{Consequences}
\newtheorem{stat}[satz]{The Statistical Hypothesis}
\newtheorem{random}[satz]{The Random Graph Idea}
\newtheorem{variant}[satz]{A Variant Approach}
\newtheorem{res}[satz]{R\'esum\'e}
\newtheorem{points}[satz]{The Qualitative Picture}
\newtheorem{conj}[satz]{Conjecture}

\newcommand{\mcal}{\mathcal}

\newcommand{\tit}{\textit}

\newcommand{\Z}{\mathbb{Z}}

\begin{document}
\begin{center}
\vspace*{1.0cm}

{\LARGE{\bf Space-Time as an Orderparameter Manifold in Random
    Networks\\ and the Emergence of Physical Points}}

\vskip 1.5cm

{\large {\bf Manfred Requardt}}\\email: requardt@theorie.physik.uni-goettingen.de 

\vskip 0.5 cm 

Institut f\"ur Theoretische Physik \\ 
Universit\"at G\"ottingen \\ 
Bunsenstrasse 9 \\ 
37073 G\"ottingen \quad Germany

\end{center}

\vspace{1 cm}

\begin{abstract}
In the following we are going to describe how macroscopic space-time
is supposed to emerge as an orderparameter manifold or superstructure
floating in a stochastic discrete network structure. As in preceeding
work (mentioned below), our analysis is based on the working
philosophy that both physics and the corresponding mathematics have to
be genuinely discrete on the primordial (Planck scale) level. This
strategy is concretely implemented in the form of \tit{cellular networks}
and \tit{random graphs}. One of our main themes is the development of
the concept of \tit{physical (proto)points} as densely entangled
subcomplexes of the network and their respective web, establishing
something like \tit{(proto)causality}. It max perhaps be said that
certain parts of our programme are realisations of some old and
qualitative ideas of Menger and more recent ones sketched by Smolin a
couple of years ago. We briefly indicate how this
\tit{two-story-concept of space-time} can be used to encode the (at
least in our view) existing non-local aspects of quantum theory
without violating macroscopic space-time causality!

\end{abstract} \newpage

\section{Introduction}
In a couple of recent papers (\cite{1} to \cite{4}) we began to
develop some facets of an extensive programme we formulated there,
i.e. rebuild ordinary continuum physics or mathematics as kind of a
coarse grained limit of a much more primordial and genuinely discrete
\tit{``theory of everything''}, including, in particular, a discrete
foundational theory of (proto) space-time as the universal receptacle
or substratum of all physical processes.

A corresponding philosophy is presently hold by a substantial minority
of workers in the diversified field described a little bit vaguely by the
catchword \tit{quantum gravity} and we commented on some of the
various approaches, at least as far as we are aware of them, in the
foregoing papers. We therefore refer the interested reader to these
papers as to references we do not mention in the following just for
sake of brevity. As an exception we mention only the early and
prophetic remarks made by Penrose in e.g. \cite{5} about the surmised
combinatorial substratum underlying our continuous space-time , the
ideas of Smolin, sketched at the end of \cite{6}, because they are
surprisingly close to our working philosophy and the work of 't Hooft
(\cite{7}) which is based on the model system of \tit{cellular
  automata} (the more rigid and regular relatives of our \tit{dynamic
  cellular networks} introduced in the following).

Our personal working philosophy is that space-time at the very bottom
(i.e. near or below the notorious Planck scale) resembles or can be
modeled as an evolving information processing cellular network,
consisting of elementary modules (with a, typically, simple internal
discrete state space) interacting with each other via dynamical bonds
which transfer the elementary pieces of information among the nodes.
It is a crucial and perhaps characteristic extra ingredient of our
framework that the bonds (i.e. the elementary interactions) are not
simply dynamical degrees of freedom (as with the nodes their internal
state space is assumed to be simple) but can a fortiori, depending on
the state of the local network environment, be created or annihilated,
i.e. can be temporarily dead or alive respectively active or inactive!
This property opens the door to the wide field of \tit{dimensional} or
 \tit{geometrical phase transitions} and \tit{selforganisation} of
space-time as an \tit{emergent} collective \tit{order parameter
  field}, floating over a much more erratic and chaotic, but largely
hidden, ``underworld''.

In this context various fundamental questions of principle pose themselves
both with respect to physics and the appropriate mathematical
apparatus. In paper \cite{1} we dealt primarily with dimensional
concepts on such discrete and irregular spaces. It furthermore became
aparent that there exist close ties to the theory of \tit{fractal
  sets}. Papers \cite{3} and \cite{4}, on the other hand, are, among
other things, devoted to the developement of several versions of
\tit{discrete analysis} and \tit{discrete differential geometry}
respectively \tit{discrete functional analysis} with certain zones of
contact with \tit{noncommutative geometry}.

All these papers were to a large extent dominated by mathematical
questions, i.e. the development of the appropriate mathematical arsenal in
order to establish something like \tit{discrete analysis and geometry}
on highly erratic spaces like our irregular (almost random)
networks. In the following we want to deal with foundational problems
of a (slightly) more physical character which, traditionally, belong more to
the realm of \tit{quantum gravity}.

As has been beautifully reviewed by Isham in various papers (see e.g.
\cite{8}) one could, among several possible attitudes, adopt the
perhaps most radical working philosophy in quantum gravity and
speculate that both \tit{quantum theory} and \tit{gravity} are merely
secondary and derived effects or, expressed in more physical terms,
so-called \tit{effective theories} of an underlying more primordial
and all embracing theory of a markedly \tit{combinatorial} flavor. A
theory comprising quantum theory and gravitation as \tit{emergent}
subtheories should, first of all, provide a framework in which the
emergence of something we are used to call \tit{space-time}
respectively \tit{quantum vacuum} can be expressed or discussed, most
notably the emergence of the \tit{continuum} from the \tit{discrete}
and the concept of physical \tit{space-time points} and their
intricate dynamical web. In other words, the main theme of this paper
will be the description of the fine structure of the substratum
underlying our continuum space-time and its dynamics together with an
investigation of its propensity for \tit{pattern creation}, that is,
patterns which, we hope, will serve as the \tit{protoforms} of the
building blocks and concepts of our present day versions of continuum
(quantum) physics and
gravity.\\[0.5cm]
Remark: We were recently kindly informed by El Naschie that a very
early source where a couple of related ideas can be found (in a
however heuristic form) is the contribution of Menger (perhaps better
known from his research on topological dimension or fractal sets like
the Menger sponge) in \cite{Me}. In this essay he entertains very
interesting ideas about the necessity of a new \tit{geometry of the
  microcosmos} based on the \tit{geometry of lumps} and the concept of
a \tit{statistical metric space}. We note in passing that these are
exactly the concepts we will develop in the following. Quite
remarkable in this respect is also Einsteins thoughtful commentary at the end
of the same volume about the \tit{discrete} and the \tit{continuum}.

In this context we also want to mention the \tit{Cantorian space-time
  approach} of El Naschie et al. who tries to model microspace as a
particular type of \tit{(random) fractal} (see e.g. \cite{Naschie}).
Another interesting early source with a possible bearing on our
approach may also be v.Neumanns concept of `\tit{continuum geometry}'
which is briefly described in \cite{Neumann}. As we became aware of
these ideas only very recently, possible connections to our strand of
ideas shall be analyzed elsewhere.\vspace{0.5cm}

A perhaps not so obvious source of inspiration should also be mentioned here
which, at first glance, is concerned with a seemingly different aspect
of nature, i.e. the science of \tit{complexity and emergent behavior}
as it is understood by e.g. the \tit{Santa Fe group} and which may be
encoded in the metapher ``\tit{complexity at the edge of chaos}'' (a
popular but beautiful account of this philosophy is \cite{9}; for
more details we recommend the bulletins issued by the institute, quite
a few corresponding references can e.g. be found in \cite{9}).

One of the reasons of our affinity to this philosophy is our belief
that many of the processes and organizing principles which are acting
behind the scene in, say, neural network science or the
\tit{coevolving fitness landscapes of complex adaptive systems} do
have their (possibly rudimentary) counterparts already on the most
primordial level of natural science i.e. fundamental physics. In a
sense this is another version of \tit{`grand unification'}.

As will be seen in the following, there exist even more surprising
links to other seemingly remote areas of current research as we
learnt very recently, a catchword being `\tit{small-world networks}'
(see \cite{10}). Some of the mechanisms being effective in our
cellular network environment are expected to drive it towards a
\tit{phase transition threshold} where space-time as kind of a
\tit{superstructure} begins to emerge exhibiting features which are
also observed in this so-called small-world scenario.

The paper is organised as follows. In the following section we define
the conceptual context. In section 3 we introduce the fundamental
framework of random graphs and prove a couple of useful
mathematical theorems about them. In section 4 we briefly describe the
presumed primordial initial phase of our network. Section 5 (together
with section 3) is the central section of the paper. In it we make the
physical concepts mathematically precise, i.e. map them on the
corresponding mathematical objects, and derive a variety of (rigorous)
results about the conjectured orderparameter manifold $ST$ and its
building blocks, the socalled \tit{`physical points'}.

\section{The Cellular Network Environment}
In this section we will set the stage for the
investigations to follow in the sections below. As was already done
in \cite{1} to \cite{4} (to which the reader is referred for more
motivation) we design the underlying substratum of our world or, more
specifically, of or physical space-time vacuum as a \tit{cellular
  network}, denoted by $QX$ (quantum space; which is, however, at the
moment only a metapher), its elementary building blocks being
\tit{nodes} and \tit{bonds}. As we have said in the introduction, the
bond interactions are not only allowed to vary in strength but, a fortiori,
can be switched off or on, depending on the state of their local 
environment. That is, stated in physical terms, bonds can be 
created or annihilated in the course of network evolution, which
(hopefully) enables the system to undergo \tit{`geometric phase
  transitions'} being accompanied by an  \tit{`unfolding'} and
\tit{`pattern formation'}, starting e.g. from a less structured
chaotic initial phase. In other words and in contrast to, say,
\tit{`cellular automata'}, which are relatively rigid and regular in
their wiring and geometric structure (in particular: with the bonds
typically being non-dynamical), our cellular networks do not carry such
a rigid overall order as an external constraint (e.g. a regular
lattice structure); their wiring is dynamical and thus behaves randomly
to some extent. To put it briefly, order and modes of regularity are
hoped to emerge via a process of \tit{`self-organization'} in a
\tit{background independent} way.\\[0.5cm]
Remarks:\begin{enumerate}
\item Some clarifying comments are perhaps in order at this point. The
  modelling of the depth structure of space-time as a cellular network
  consisting of nodes and bonds should not necessarily be understood
  in a plain bodily sense. One should rather consider it as a
  representation or emulation of the main characteristics of the
  physical scenario, as it is common practice in the art of model building
  being prevalent in modern physics. There may, in particular, exist a
  variety of superficially different systems, the logical structure of
  which can nevertheless be encoded in roughly the same abstract
  underlying network model. It is our personal belief that such a
  discrete network, governed by a relatively simple but cleverly
  chosen dynamical law, is capable of generating most if not all of
  the phenomena and emergent laws on which our ordinary continuum
  physics is usually grounded. That such a hypothesis is not entirely
  far-fetched may e.g. be inferred from the emerging complexity of such
  a simple cellular automaton model as the famous \tit{`game of
    life'}.
\item A typical example is the \tit{geometry of lumps} envisaged by
  Menger. Take as lumps the hypothetical \tit{`infinitesimal'} grains
  of space or space-time which cannot be further resolved (be it in a
  practical or principal sense). Let them overlap according to a
  certain rule so that they can interact or exchange information. Draw
  a node for each such lump and a bond for each two lumps which happen
  to overlap. In \tit{combinatorial topology} such a
  \tit{combinatorial complex} is called the \tit{nerve} of the
  \tit{set system} (cf. e.g. \cite{Hopf}). In a next step one may
  encode the respective strengths of interaction or degrees of overlap
  in a valuation of the corresponding bonds, yielding a cellular
  network of the kind we are having in mind. A fortiori one can make
  these mutual overlaps into dynamical variables, i.e. let them change
  in ``time''.
\end{enumerate}
\vspace{0.5cm}

A certain class of relatively simple cellular networks is the following.
\begin{defi}[Class of Cellular Networks] \hfill  
\begin{enumerate}
\item ``Geometrically'' our networks represent at each fixed
{\em `clock time'} are {\em `labeled graphs'}, i.e. they consist of nodes
\{$n_i$\} and bonds \{$b_{ik}$\}, with the bond $b_{ik}$ connecting
the nodes (cells) $n_i$, $n_k$. We assume that the graph has neither
elementary loops nor multi-bonds, that is, only nodes with $i\neq k$
are connected by at most one bond.
\item At each site $n_i$ we have a local node state $s_i\in
q\cdot\Z$ with $q$, for the time being, a certain not further
specified  elementary quantum. The bond variables $J_{ik}$, attached
to $b_{ik}$, are in the most simplest cases assumed to be two- or
three-valued, i.e. $J_{ik}\in \{\pm 1\}\quad\mbox{or}\quad J_{ik}\in \{\pm
  1,0\}$
\item As to the (time (in)dependent) distribution of bonds over the graph there
  exist several (physically motivated) possibilities as will be
  explained in the following. In defining the mathematical model one
  may choose to start with an a priori fixed set of bonds
  (i.e. independent of e.g. \tit{`clock time'} or something else) and
  shift the corresponding dynamics to the bond variables
  $J_{ik}$. Physically one may, on the other hand, prefer to make this set of bonds itself
  a dynamical set, i.e. with bonds vanishing and emerging. Both
  schemes have their respective advantages.
\end{enumerate}
\end{defi}

Remarks \hfill 
\begin{enumerate}
\item In the proper graph context the notions {\em `vertex'} and 
{\em `edge'} are perhaps more common (see e.g. \cite{11}). As to other concepts occurring in graph theory see below.
\item This is, in some sense, the simplest choice one can
make. It is an easy matter to employ instead more complicated or
reduced internal
state spaces like, say, groups, manifolds etc. One could in particular
replace $\Z$ by one of its quotient groups as e.g. $\Z_n$ or impose suitable  boundary 
conditions.
\item It is useful to give the bonds $b_{ik}$ an 
{\em `orientation'}, i.e. (understood in an precise
algebraic/geometric sense) $b_{ik}=-b_{ki}$.  
This implies the compatibility conditions $J_{ik}=-J_{ki}$.
\end{enumerate}
\vspace{0.5cm}

In a next step we impose a dynamical law on our model
network. In this respect we are of course inspired by {\em `cellular
  automaton laws'} (see e.g. \cite{12}). The main difference is however
that in our context also the bonds are dynamical degrees of freedom
and that, a fortiori, they can become dead or alive (active or
inactive), so that the whole net is capable of performing drastic 
topological/geometrical changes in the course of clock time.

A large class of dynamical `{\it local laws}' are e.g. the following
ones: We assume that all the nodes/bonds at discrete `{\it (clock)
  time}' $t+\tau$, $\tau$ an elementary clock time step (i.e.
$t=z\cdot\tau$ with $z\in \Z$), are updated according to a certain
local rule which relates the internal node space/bond space of each
node/bond at time $t+\tau$ with the states of the nodes/bonds of a
certain fixed local neighborhood at time $t$.

It is important that, generically, such a law does not lead to a
reversible time evolution, i.e. there will typically exist attractors
in total phase space (the overall configuration space of
the node and bond states). On the other hand, there exist strategies
to develop reversible network laws.

A crucial ingredient of our network laws is what we would like to call
a `{\it hysteresis interval}'. We will assume that our network, $QX$,
starts from a densely entangled `{\it initial phase}' $QX_0$, in which
practically every pair of nodes is on average (to be understood in a
statistical sense) connected by an `{\it active}' bond, i.e.
$J_{ik}=\pm1$.  Our dynamical law will have a built-in mechanism which
switches bonds off (more properly: sets $J_{ik}=0$) if local
fluctuations among the node states become too large. Then there is a
certain hope that this mechanism may trigger an `{\it unfolding phase
  transition}', starting from a local seed of spontaneous large
fluctuations towards a new phase (an attractor) carrying an emergent
`{\it super structure (proto space-time)}', which we want to identify
with the hidden discrete substratum of our ordinary continuous space-time.

One example of such a law is given in the following definition.
\begin{defi}[Example of a Local Law]
At each clock time step a certain `{\em quantum}' $q$
is exchanged between, say, the nodes $n_i$, $n_k$, connected by the
bond $b_{ik}$ such that 
\begin{equation} s_i(t+\tau)-s_i(t)=q\cdot\sum_k
  J_{ki}(t)\end{equation}
(i.e. if $J_{ki}=+1 $ a quantum $q$ flows from $n_k$ to $n_i$ etc.)\\
The second part of the law describes the {\em back reaction} on the bonds
(and is, typically, more subtle). This is the place where the
so-called `{\em hysteresis interval}' enters the stage. We assume the
existence of two `{\em critical parameters}'
$0\leq\lambda_1\leq\lambda_2$ with:
\begin{equation} J_{ik}(t+\tau)=0\quad\mbox{if}\quad
  |s_i(t)-s_k(t)|=:|s_{ik}(t)|>\lambda_2\end{equation}
\begin{equation} J_{ik}(t+\tau)=\pm1\quad\mbox{if}\quad 0<\pm
  s_{ik}(t)<\lambda_1\end{equation}
with the special proviso that
\begin{equation} J_{ik}(t+\tau)=J_{ik}(t)\quad\mbox{if}\quad s_{ik}(t)=0
\end{equation}
On the other side
\begin{equation} J_{ik}(t+\tau)= \left\{\begin{array}{ll} 
\pm1 & \quad J_{ik}(t)\neq 0 \\
0    & \quad J_{ik}(t)=0
\end{array} \right. \quad\mbox{if}\quad
\lambda_1\leq\pm
  s_{ik}(t)\leq\lambda_2 
\end{equation}
In other words, bonds are switched off if local spatial charge
fluctuations are too large, switched on again if they are too
small, their orientation following the sign of local charge
differences, or remain inactive.
\end{defi}
Remarks \hfill 
\begin{enumerate}
\item We do not choose the ``current'' $q\cdot
J_{ik}$ proportional to the ``voltage difference'' $(s_i-s_k)$ as e.g.
in Ohm's law since we favor a {\it non-linear} network which is
capable of {\it self-excitation} and {\it self-organization} rather than
{\it self-regulation} around a relatively uninteresting equilibrium state!
The balance between dissipation and amplification of spontaneous
fluctuations has
however to be carefully chosen (``{\it complexity at the edge of chaos}'')
\item We have emulated these local network laws on a computer and made
  extensive numerical studies of a lot of network or graph characteristics
  (which will be published elsewhere). It is not yet clear whether
  the above simple network law does already everything we are expecting. We
  studied networks up to roughly five thousand nodes. This is still
  quite a small number if one intends to simulate the whole universe
  where node numbers of $10^{100}$ or so should be envisaged.
  Furthermore, the investigation of certain complex graph properties is
  almost \tit{non-polynomially complete}.  In any case, it is
  fascinating to observe the enormous capability of such intelligent
  networks to find attractors very rapidly, given the enormous
  accessible phase space (as to this particular and important feature
  confer the remarks of S.Kauffman in \cite{9})
\item In the above class of laws a direct bond-bond interaction is not
  yet implemented. We are prepared to incorporate such a (possibly
  important) contribution in a next step if it turns out to be
  necessary. In any case there are not so many ways to do this in a
  sensible way. Stated differently, the class of possible physically
  sensible interactions is perhaps not so numerous.
\item Note that -- in contrast to e.g. Euclidean lattice field theory --
the so-called {\it `clock time'} $t$ is, for the time being, not
standing on the same footing as potential ``coordinates'' in the
network (e.g.  curves of nodes/bonds). We rather expect so-called
{\it `physical time'} to emerge as sort of a secondary collective
variable in the network, i.e. to be different from the clock time
(while being of course functionally related to it). \label{no4}
\end{enumerate}
\vspace{0.5cm}

We nevertheless regard Remark $4$ to be consistent with the spirit of
relativity. What Einstein was really teaching us is that there is a
(dynamical) interdependence between what we experience as space
respectively time, not that they are absolutely identical! In any
case, the assumption of an overall clock time is at the moment only
made just for convenience in order to make the model system not too
complicated. If our understanding of the complex behavior of the
network dynamics increases, this assumption may be weakened in favor
of a possibly local and/or dynamical clock frequency. A similar
attitude should be adopted concerning concepts like
\tit{`Lorentz-(in)covariance'} which we also consider as
\tit{`emergent'} properties. It is needless to say that it is of
tantamount importance to understand the way how these patterns do
emerge from the relatively chaotic background, questions which will be
addressed in future work (as to some related ideas about such issues
see also \cite{Kauffman} or \cite{Sorkin}).

The above example of a network law is only one candidate from a
whole class of such laws. For one, it is quite evident that the
`\tit{local state spaces}' living over the respective nodes and bonds
can be chosen in a more general way. For another, the local dynamical
law can be chosen more general.
\begin{defi}[General Local Law on Cellular Networks]
Each node
$n_i$ can be in a number of internal states $s_i\in \cal S$. Each
bond $b_{ik}$ carries a corresponding bond state $J_{ik}\in\cal J$. Then
the following general transition law is assumed to hold:
\begin{equation} s_i(t+\tau)=ll_s(\{s_k(t)\},\{J_{kl}(t)\})
\end{equation}
\begin{equation} J_{ik}(t+\tau)=ll_J(\{s_l(t)\},\{J_{lm}(t)\})
\end{equation}
\begin{equation}
(\underline{S},\underline{J})(t+\tau)=LL((\underline{S},\underline{J})(
t))
\end{equation}
where $ll_s$, $ll_J$ are two maps (being the same over the whole
graph) from the state space of a local
neighborhood of the node or bond on the lhs 
to $\cal S, J$, yielding the
updated values of $s_i$ and $J_{ik}$. $\underline{S}$ and
$\underline{J}$ denote the global states of the nodes and bonds and
$LL$ the global law built from the local laws at each node or bond.
\end{defi}
We close this section with a brief r\'esum\'{e} of the characteristics
an interesting network dynamics should encode.
\begin{res}
  Irrespectively of the technical details of the dynamical evolution
  law under discussion it should emulate the following, in our view
  crucial, principles, in order to match certain fundamental
  requirements concerning the capability of `{\it emergent}' and `{\it
    complex}' behavior.
\begin{enumerate}
\item As is the case with, say, gauge theory or general relativity,
  our evolution law on the surmised primordial level should implement
  the mutual interaction of two fundamental substructures, put
  sloppily: ``{\em geometry}'' acting on ``{\em matter}'' and vice
  versa, where in our context ``{\em geometry}'' is assumed to
  correspond in a loose sense with the local and/or global bond states
  and ``{\em matter}'' with the structure of the node states.
\item By the same token the alluded {\em selfreferential} dynamical
  circuitry of mutual interactions is expected to favor a kind of
  `{\em undulating behavior}' or `{\em selfexcitation}' above a return
  to some uninteresting `{\em equilibrium state}' as is frequently
  the case in systems consisting of a single component which directly
  acts back on itself. This propensity for the `{\em autonomous}'
  generation of undulation patterns is in our view an essential
  prerequisite for some form of ``{\em protoquantum behavior}'' we
  hope to recover on some coarse grained and less primordial level of
  the network dynamics.
\item In the same sense we expect the overall pattern of switched-on and
 -off bonds to generate a kind of ``{\em protogravity}''.
\end{enumerate}
\end{res}
Remark: As in the definition of evolution laws of `\tit{spin
  networks}' by e.g. Markopoulou, Smolin and Borissov (see \cite{13} or
\cite{14}), there are in our case more or less two possibilities:
treating evolution laws within an integrated space-time formalism or
regard the network as representing space alone with the time evolution
being implanted via some extra principle ( which is the way we have
chosen above). As the interrelation of these various approaches and
frameworks is both very interesting and presently far from obvious we
plan to compare them elsewhere.

\section{The Cellular Network as a (Random) Graph}
There are many different aspects one can study as regards our class of
complex cellular networks. As in a purely geometric sense they are
\tit{graphs} it is, in a first step, a sensible strategy to supress
all the other features like e.g. the details of the internal state
spaces of nodes and bonds and concentrate instead on its pure
\tit{`wiring diagram'} and its reduced dynamics. This is already an
interesting characteristic of the network (perhaps somewhat
reminiscent of the \tit{`Poincar\'e map'} in the theory of chaotic
systems) as bonds can be switched on and off in the course of time so
that already the wiring diagram will constantly change. Furthermore,
as we will see, it encodes the complete \tit{near-} and \tit{far-order
  structure} of the network, that is, it tells us which regions are
experienced as near by or far apart (in a couple of possible physical
ways such as strength of correlations, with respect to some physical metric or
\tit{`statistical distance'}). Evidently this is one of the crucial
features we expect from something like physical space-time.

We start with the introduction of some graph theoretical definitions.
\begin{defi}[Simple Locally Finite (Un)directed Graph)]\hfill
\begin{enumerate}
\item We write the `{\em simple}' `{\em labeled}' graph as $G:=(V,E)$ where $V$ is the
  countable set of {\em nodes} $\{n_i\}$ (or {\em vertices}) and $E$ the set of
  {\em bonds} ({\em edges}). The graph is called {\em simple} if there do not exist
  elementary `{\em loops}' and `{\em multiple edges}', in other words: each
  existing bond connects two different nodes and there exists at most
  one bond between two nodes. (We could of course also discuss more
  general graphs). Furthermore, for simplicity, we assume the graph to
  be {\em connected}, i.e. two arbitrary nodes can be connected by a
  sequence of consecutive bonds called an {\em edge sequence} or `{\em
    walk}'. A
  minimal edge sequence, that is one with each intermediate node
  occurring only once, is called a `{\em path}' (note that these definitions
  may change from author to author).
\item We assume the graph to be `{\em locally finite}' (but this not
  always really necessary), that is, each node is incident with only a
  finite number of bonds. Sometimes it is useful to make the stronger
  assumption that this `{\em vertex degree}', $v_i$, (number of bonds
  being incident with $n_i$), is globally bounded away from $\infty$.
\end{enumerate}
\end{defi}
\begin{obdef}
  Among the paths connecting two arbitrary nodes there exists at least
  one with minimal length. This length we denote by
  $d(n_i,n_k)$. This d has the properties of a metric, i.e:
\begin{eqnarray} d(n_i,n_i) & = & 0\\ d(n_i,n_k) & = &
d(n_k,n_i)  \\d(n_i,n_l) & \leq & d(n_i,n_k)+d(n_k,n_l) \end{eqnarray}
\end{obdef}
(The proof is more or less evident).
\begin{koro}
With the help of the metric one gets a natural
neighborhood structure around any given node, where ${\cal U}_m(n_i)$
comprises all the nodes, $n_k$, with $d(n_i,n_k)\leq m$, 
$\partial{\cal U}_m(n_i)$
the nodes with $d(n_i,n_k)=m$.
\end{koro}
Remark: The restriction to connected graphs is, for the time being,
only made for convenience. If one wants to study geometric phase
transitons of a more fragmented type, it is easy to include these more
general types of graphs. In the context of \tit{random graphs} ( which
we will introduce below) one can even derive probabilistic criteria
concerning geometric properties like connectedness etc.
\begin{ob}[Projected Graph Process]
The {\em projection} of the full network process on the space of
corresponding graphs over, say, a given fixed set of nodes defines
what one may call a {\em projected graph process} with respect to {\em
  clocktime $t$}. The reduced dynamics or evolution of the graph,
$G(t)$, i.e. the wiring diagram of the network, is then expressed by
the creation and annihilation of bonds during consecutive time steps.

Note however that this type of process is, in general, not an
intrinsic one. The evolution law on the graph level is only defined
extrinsically via the network law and generically cannot be reduced to
a law acting on graph space as the projection is accompanied by an
information loss. That is, different network states may typically
project on the same graph state and, a fortiori, will lead in most
cases after another time step even to different graph states.
\end{ob}

The graphs or networks we are actually interested in are expected to
be extremely large. According to our philosophy they are to emulate
the full physical vaccum together with all its more or less
macroscopic excitations, or, in other words, the entire evolving
universe. Furthermore the assumed \tit{clocktime interval} $\tau$ is
extremely short (in fact of Plancktime  order). On the other side, it
is part of our working philosophy that the phenomena we are observing
in e.g. present day high energy physics and, a fortiori, macroscopic
physics are of the nature of collective (frequently large scale)
excitations of this medium both with respect to space and time (in
Planck units). In other words, each of these patterns is expected to
contain, typically, a huge amount of nodes and bonds and to stretch
over a large number of clocktime intervals. This then suggests the
following approach which has been fruitful again and again in modern
physics.
\begin{stat}
Following the above arguments it makes sense to study so-called `{\em
  graph properties}' within a certain {\em statistical framework} to be
further explained below.
\end{stat}

There are some remarks in order as this concept has at least two, to
some extent different, facets. \\[0.5cm]
Remarks:
\begin{enumerate}
\item First, there exists the usual strategy to average over small
  scale events or degrees of freedom in order to aquire some (more or
  less) coarse grained but, on the other side, more generic
  information about the evolving network or graph (examples being
  given below).
\item Second, while our network is, on the one side, up to now
  completely deterministic, one may nevertheless entertain the idea
  that, irrespective of the details of potential initial conditions
  or of our network law, there exist at least whole {\em classes} of
  them which lead to the same gross phenomena (whereas the underlying
  microstates at each given time step may be quite distinct). It hence
  seems to be advisable to group appropriate sets of microstates
  into entities of a slightly higher order of organisation called
  \tit{phases}. A typical \tit{initial phase} will be decribed below.
\item A characteristic phenomenon in this context is the
  following. Irrespective of the huge accessible \tit{microscopic
    phase space}, networks like ours typically manage to reach an
  \tit{attractor} (which may of course depend on the initial phase)
  after a frequently surprisingly short \tit{transient} which makes
  the assumed \tit{robustness} against the details of initial
  conditions quite apparent. Similar strategies have been adopted in
  the investigation of cellular automata (see e.g. \cite{Wolf} or \cite{Farm}).
\item Another strand of reasoning is based on a slightly different
  bundle of ideas. Consider all possible graphs, $\{G_i\}$, over $n$
  vertices or nodes. In contrast to the above approach, which deals
  rather with `\tit{practical statistics}', we will now construct a
  \tit{probability space} in at least two ways. 
\end{enumerate}    
\begin{random}Take all possible labeled graphs over $n$ nodes as
  probability space $\cal{G}$ (i.e. each graph represents an
  elementary event). The maximal possible number of bonds is
  $N:=\binom{n}{2}$, which corresponds to the unique {\em simplex
    graph} (denoted usually by $K_n$). Give each bond the {\em
    independent probability} $0\leq p\leq 1$, (more precisely, $p$ is
  the probability that there is a bond between the two nodes under
  discussion). Let $G_m$ be a graph over the above vertex set having
  $m$ bonds. Its probability is then
\begin{equation}pr(G_m)=p^m\cdot q^{N-m}\end{equation}
where $q:=1-p$. There exist ${N\choose m}$ different labeled
$m$-graphs $G_m$ and the above probability is correctly normalized,
i.e.
\begin{equation}pr({\cal G})=\sum_{m=0}^N {N\choose m}p^mq^{N-m}=(p+q)^N=1\end{equation}
This probability space is sometimes called the space of `{\em
  binomially random graphs}' and denoted by ${\cal G}(n,p)$. Note that
the number of edges is binomially distributed, i.e.
\begin{equation}pr(m)=\binom{N}{m}p^mq^{N-m}\end{equation}
and
\begin{equation}\langle m\rangle=\sum m\cdot pr(m)=N\cdot p\end{equation}
\end{random}
Proof of the latter statement:
\begin{equation}\langle
  m\rangle=d/d\lambda|_{\lambda=1}\left(\sum\binom{N}{m}(\lambda
    p)^mq^{N-m}\right)=d/d\lambda|_{\lambda=1}(\lambda
  p+q)^N=Np\end{equation}
or, with $e_i$ being the independent `\tit{Bernoulli
  $(0,1)$-variables}' (as to this notion cf. e.g. \cite{Feller}) belonging to the bonds:
\begin{equation}\langle e_i\rangle=p\quad\text{hence}\quad\langle
  m\rangle=\sum_1^N\langle e_i\rangle=Np \end{equation}
$\Box$\hfill\\
(The use of the above Bernoulli variables leads also to some
conceptual clarifications in other calculations). 
\begin{variant}A slightly different probability space can be
  constructed by considering only graphs with a fixed number, $m$, of
  bonds and give each the probability ${N\choose m}^{-1}$ as there are
  exactly ${N\choose m}$ of them. The corresponding probability space,
  ${\cal G}(n,m)$, is called the space of `{\em uniform random graphs}'.
\end{variant}

The latter version is perhaps a little bit more common in pure mathematics
as this concept was introduced  mainly for purely combinatorial
reasons which have nothing to do with our own strand of ideas. The
whole theory was rather developed by Erd\"os and R\'enyi in the late
fifties and early sixties to cope with certain notorious (existence)
problems in graph theory (for more information see e.g. \cite{Random}
and \cite{Combi}, brief but concise accounts can also be found in
chapt.VII of \cite{11} or \cite{Kar}).
\begin{ob}The two random graph models behave similarly if $m\approx
  p\cdot N$. Note however, that there exists a subtle difference
  between the two models anyway. In the former model all elementary
  bond random variables are {\em independent} while in the latter case
  they are (typically weakly) dependent.
\end{ob}
(While being plausible this statement needs a proof which can be found
in e.g. \cite{Random}).\vspace{0.5cm}

The really fundamental observation made already by Erd\"os and R\'enyi (a
rigorous proof of this deep result can e.g. be found in \cite{Combi})
is that there are what physicists would call \tit{phase transitions}
in these \tit{random graphs}. To go a little bit more into the details
we have to introduce some more graph concepts.
\begin{defi}[Graph Properties]{\em Graph properties} are certain
  particular {\em random
    variables} (indicator functions of so-called events) on the above
  probability space ${\cal G}$. I.e., a graph property, $Q$, is
  represented by the subset of graphs of the sample space having the
  property under discussion. To give some
  examples: i) connectedness of the graph, ii) existence and number of certain particular subgraphs (such as
  subsimplices etc.), iii) other geometric or topological graph
  properties etc.
\end{defi}
Remark: In addition to that there are other more general random
variables (\tit{`graph characteristics'}) describing the fine
structure of graphs, some of which we will introduce
below.\vspace{0.5cm}

In this context Erd\"os and R\'enyi made the following important
observation.
\begin{ob}[Threshold Function]A large class of {\em graph properties}
  (e.g. the `{\em monotone increasing ones}', cf. the above cited
  literature) have a so-called `{\em threshold function}', $m^*(n)$,
  so that for $n\to\infty$ the graphs under discussion have {\em
    property} $Q$ {\em almost shurely} for $m(n)>m^*(n)$ and {\em
    almost shurely not} for $m(n)<m^*(n)$ or vice versa (more
  precisely: for $m(n)/m^*(n)\to \infty\;\text{or}\;0$; for the
  details see the above literature). The above version applies to the
  second kind of graph probability space, ${\cal G}(n,m)$. A
  corresponding result holds for ${\cal G}(n,p)$ with $p(n)$ replacing
  $m(n)$. That is, by turning on the probability $p$, one can drive
  the graph one is interested in beyond the phase transition threshold
  belonging to the graph property under study. Note that, by
  definition, threshold functions are only unique up to
  ``factorization'', i.e. $m^*_2(n)= O(m^*_1(n))$ is also a
  threshold function.
\end{ob}
\begin{bsp}[Connectedness]The threshold function for the graph
  property {\em connectedness} is 
\begin{equation}m^*(n)=n/2\cdot\log(n)\;\text{respectively}\;
  p^*(n)=\log(n)/n\end{equation}
Note that with the help of the above observation, i.e. for $m\approx
p\cdot\binom{n}{2}$, we have for $n$ large: $\binom{n}{2}\approx
n^2/2$ and hence $p\cdot n^2/2\approx n\log(n)$, i.e. $p\approx \log(n)/n$
\end{bsp}
\begin{obdef}[Combinatorial Graph Process]
  It is perhaps conceptually more intuitive to envisage the passage of
  a random graph through its various epochs in a more dynamical way by
  introducing a {\em combinatorial graph process}, starting from the
  {\em empty graph} and adding at each discrete ``time step'' one
  edge, chosen at random. Each element $\tilde{G}$ consists of a
  sequence of random graphs $\{G_m\}_0^N$. Note however that no real
  dynamics in e.g. a physical sense is implied. Therefore this kind of
  combinatorial process should not be confused with the real dynamics
  and evolution of our networks and graphs.
\end{obdef}

In the following our main thrust will go towards the developement of
the concept of `\tit{proto space-time}' as an `\tit{order parameter
  manifold}' or `\tit{superstructure}' floating in our network $QX$
and the concept of `\tit{physical points}' together with their
respective `\tit{near- and far-order}'. We therefore consider it
worthwhile to illustrate the method of random graphs, graph
properties and graph characteristics by applying it to a particular feature being of importance
in the sequel.
\begin{defi}[Subsimplices and Cliques]With $G$ a given fixed graph and
  $V_i$ a subset of its vertex set $V$, the corresponding {\em induced
    subgraph} over $V_i$ is called a subsimplex (ss),
  $S_i$, or {\em complete subgraph}, if all its nodes are connected by
  a bond. In this class there exist certain {\em maximal subsimplices}
   $(mss)$, that is, every addition of another node destroys this
  property. These $mss$ are called {\em cliques} in combinatorics.
\end{defi}
For reasons to be explained below we are extremely interested in such
$mss$ and their behavior (which depends of course on the given graph
$G$ under discussion). It is sometimes a certain art to design an
appropriate random variable implementing a particular graph property
and work out the necessary combinatorial probabilities. In our
specific example we proceed as follows: We consider all possible
graphs, $G$, over the fixed vertex set $V$ of $n$ nodes. For
each subset $V_i\subset V$ of order $r$ (i.e. number of elements) we
define the following random variable:
\begin{equation}X_i(G):=
\begin{cases}1 & \text{if $G_i$ is an $r$-simplex},\\  
 0 & \text{else}
\end{cases}
\end{equation}
where $G_i$ is the corresponding induced subgraph over $V_i$ with
respect to $G\in {\cal G}$ (the probability space). Another random
variable is then the \tit{number of $r$-simplices in $G$}, denoted by
$Y_r(G)$ and we have:
\begin{equation}Y_r=\sum_{i=1}^{\binom{n}{r}}X_i\end{equation}
with $\binom{n}{r}$ the number of $r$-subsets $V_i\subset V$. With respect
to the probability measure introduced above we then have for the
\tit{expectation values}:
\begin{equation}\langle Y_r \rangle = \sum_i \langle X_i \rangle\end{equation}
and
\begin{equation}\langle X_i \rangle = \sum_{G\in{\cal G}} X_i(G)\cdot
  pr(G_i=\text{$r$-simplex in}\;G).\end{equation} 
With the sum running
over all $G\in {\cal G}$ and $X_i$ being one or zero we get
\begin{equation}\langle X_i \rangle = pr(G_i\;\text{an $r$-simplex,
    $G-E_i$ an arbitrary graph})\end{equation}
where $G-E_i$ is the remaining graph after all the edges belonging to
$G_i$ have been eliminated. This yields 
\begin{equation}\langle X_i \rangle =
  p^{\binom{r}{2}}\cdot\sum_{G'\in{\cal G}'} pr(G')\end{equation}
where ${\cal G}'$ is the probability space of graphs over $V$ with all
the bonds $E_i$ being omitted. The maximal possible number of bonds
belonging to ${\cal G}'$ is
\begin{equation}|E'|=\binom{n}{2}-\binom{r}{2}.\end{equation}
Each of these bonds can be on or off with probability $p$ or
$(1-p)$. To each graph of ${\cal G}'$ belongs a unique labeled
sequence of $p$'s and $q$'s and every such sequence does occur
(i.e. with either $p$ or $q$ at label $i$). We hence have
\begin{equation}\sum_{G'}pr(G')=(p+q)^{|E'|}=1^{|E'|}=1\end{equation}
and we get
\begin{equation}\langle X_i \rangle = p^{\binom{r}{2}}\end{equation}

There are now exactly $\binom{n}{r}$ possible $r$-subsimplices over
the node set $V$, hence we arrive at the following important conclusion:
\begin{conclusion}[Subsimplices]The expectation value of the random
  variable {\em number of $r$-subsimplices} is
\begin{equation}\langle Y_r \rangle = \binom{n}{r}\cdot
  p^{\binom{r}{2}}\end{equation}
\end{conclusion}
It is remarkable, both physically and combinatorially, that this
quantity as a function of $r$, i.e. the order of subsimplices, has
quite a peculiar numerical behavior. We are, among other things,
interested in the typical order of cliques (where typical is
understood in a probabilistic sense).
\begin{obdef}[Clique Number]The maximal order of a clique contained in
  $G$ is called its {\em clique number}, $cl(G)$. It is another random
  variable on the probability space ${\cal G}(n,p)$.
\end{obdef}
An analysis of the above expression $\langle
Y_r\rangle=\binom{n}{r}\cdot p^{\binom{r}{2}}$ as a function of $r$
shows that it is typically very large (for $n$ sufficiently large) for
all $r\leq r_0$ and drops rapidly to zero for $r> r_0$ for $r_0$
some \tit{critical value} to be determined.
\begin{conclusion}From the above one can infer that this value $r_0$
  is an approximation for the above defined {\em clique number} of a
  typical random graph, depending only on $p$ and $n$. In other words,
  it approximates the order of the largest occurring clique in the
  respective graph. A good approximation of $r_0$ is
\begin{equation}r_0\approx 2\log(n)/\log(p^{-1})+ O(\log\log(n))\end{equation} 
(cf. chapt. XI.1 of \cite{Random}).
\end{conclusion}
\begin{bsp}$(n=100,\;p=1/2)\Rightarrow \;r_0\approx 12$
\end{bsp}

$r_0$ is an approximation of the typical order of the highest
occurring clique in a random graph (understood in a probabilistic
sense). For our investigation of the fine structure of space-time it
is however of tantamount importance to acquire more detailed
information about the typical number and distribution of cliques in a
random graph for varying $r$ (note that the above formula provides, up
to now, only an estimate of the number of $r$-simplices). It seems to be a
much more ambitious task to deal with the corresponding case in the
extremal situation. This problem can however be solved by
a slightly tricky approach.

The probability that the induced subgraph over $r$ arbitrary but fixed
vertices, $n_1,\ldots,n_r$, (with the graph $G$ given), is a $ss$ is
$p^{\binom{r}{2}}$ (see above). If it is, a fortiori, a clique or
$mss$, this means that the addition of a single further vertex would
destroy the property being a $ss$. In other words, each of the
vertices in the graph $G$, different from the above $n_1,\ldots,n_r$,
is connected with the above vertex set, $V_r$, by fewer than $r$
bonds. This can now be quantitatively encoded as follows:
\begin{ob}[distribution of $mss$ or cliques] The probability that the
  induced subgraph $G_r$ over $r$ arbitrarily chosen vertices is
  already a $mss$ is the product of the {\em independent}
  probabilities that it is a $ss$ and that {\em each} of the remaining
  $(n-r)$ vertices is connected with $V_r$ by fewer than $r$
  bonds. The latter probability is $(1-p^r)^{n-r}$, hence:
\begin{equation}pr(G_r\;\text{is a clique})=(1-p^r)^{n-r}\cdot
  p^{\binom{r}{2}}\end{equation}
From this we can immediately infer for $Z_r$, the number of
$r$-cliques in the random graph, the following relation
\begin{equation}\langle Z_r
  \rangle=\binom{n}{r}\cdot(1-p^r)^{n-r}\cdot
  p^{\binom{r}{2}}\end{equation}
\end{ob}
\begin{conclusion}By an estimation of the {\em variance} of $Z_r$ we
  can conclude that the typical orders of occurring cliques lie
  between $r_0(n)/2$ and $r_0(n)$.
\end{conclusion}
\begin{bem}To make the above reasoning completely rigorous, it is again helpful to
  exploit the properties of the elementary $(0,1)$-edge-variables
  $e_i$. The probability that $r$ arbitrarily selected bonds exist in
  the random graph is $pr(e_1=\ldots=e_r=1)=p^r$, the complimentary
  possibility (i.e., that some of these bonds are missing), has hence the probability $(1-p^r)$.
\end{bem}
\begin{res}[Random Graph Approach]There is of course no absolute
  guarantee that our network, following a deterministic evolution law
  and typically reaching after a certain {\em transient} one of
  possibly several attractors, can in every respect be regarded as the
  evolution of true random graphs. In other words, its behavior may
  not be entirely random. Quite to the contrary, we expect a {\em shrinking
  of phase space} during its evolution which manifests itself on a more
  macroscopic level as {\em pattern creation}.
  
  The underlying strategy is rather the following. At each clock time
  step, $G(t)$ is a graph having a definite number of bonds, say $m$.
  Surmising that $G(t)$ is sufficiently generic or typical we can
  consider it as a typical member of the corresponding family ${\cal
    G}(n,m)$ or ${\cal G}(n,p)$. Via this line of inference we are
  quite confident of being able to get some clues concerning the
  qualitative behavior of our network. As to this working philosophy
  the situation is not completely different from the state of affairs
  in many areas of, say, ordinary statistical physics. But
  nevertheless, a more detailed analysis (underpinned by concrete
  examples) of the validity of this ansatz would be desirable and
  shall be given in forthcoming work.

What we said above is on the one side corroborated by our own
extensive computer simulations (performed by our former student
Th. Nowotny). On the other side such a possibly characteristic
numerical deviation between theoretical results based on certain apriori
statistical assumptions and computer simulations of concrete models
was also found in \cite{Antonsen} by Antonsen, who devised an approach which is
different from ours in several respects but is similar in the
general spirit.
\end{res}
\section{A Qualitative Description of the Presumed Initial Phase}
The central theme of this paper is the description and analysis of a
certain superstructure, $ST$, emerging within our network $QX$ as a
consequence of a process which can be interpreted as a geometrical
phase transition. In this picture $ST$, which we experience as
space-time, plays the role of an order parameter manifold. Its
emergence signals the transition from a \tit{disordered} and
\tit{chaotic} initial phase to a phase developing a
\tit{near-/far-order}, i.e. a \tit{causal structure}, and stable
\tit{`physical points'} or \tit{`lumps'} (Menger).

As explained above, it is advantageous at various intermediate steps
to coarse grain the amount of information being buried in the
microstates of the network and concentrate rather on the geometric
information encoded, say, in its pure wiring diagram, i.e. the
corresponding graph process $G(t)$. Furthermore, as is the typical
case with dynamical systems of such an extreme complexity, we are
typically interested in the generic behavior and patterns and not so
much in every, possibly only contingent, detail of the microstates of
the network. That means, the natural object of our interest is what we
called in section 3 a \tit{phase}, that is, a whole class of
microstates behaving similarly in a certain ``macroscopic'' or
large-scale respect (yet to be defined). This is in complete analogy
with the ideas of ordinary statistical physics and hence calls for a
statistical approach and analysis of the kind we mentioned in the
preceding sections.

In a first step we have to describe the presumed \tit{`initial phase'}
of our network. This is the phase which, due to its chaotic behavior,
cannot yet support a superstructure like $ST$ and which is, a
fortiori, assumed to be free of any other stable type of
pattern. After some time of model building and experimenting emerged
as the, in our view, most natural and plausible candidate a phase
where, on average, almost every node is connected with any other node
so that, among other things, all the nodes are roughly staying on equal
footing with respect to each other. On the other side, the expected inherent
wild fluctuations in the network entail that bonds happen to be
permanently created and annihilated.\\[0.5cm]
Remarks: There exist several reasons for selecting this scenario as
initial phase.\\
\begin{enumerate}
\item For one, we want to emulate the transition of the universe from
  chaos to \tit{complex} order. We expect that a phase in which
  practically all the nodes and bonds are standing on an equal footing
  with respect to each other and in which ``everything interacts with
  everything'' in a highly erratic manner (given the presumed
  extremely high vertex degree) guarantees the complete absence of
  stable patterns or order of any other kind.
\item For another, our approach (as kind of a \tit{``theory of
    everything''}) shall also be capable of describing the \tit{``big
    bang era''} of our universe, i.e. the unfolding of space-time,
  starting from a tiny nucleus. It will turn out that such a scenario
  can in fact be described in a very natural and \tit{``unforced''}
  way within our network framework which is of course very satisfying.
\item Lastly, such an initial phase is, as far as we can see, almost
  the only one which is free of contingent or adhoc assumptions like

  e.g. a particular choice of the initial distribution of active bonds
  etc.
\end{enumerate}
\begin{assumption}[Initial Conditions] We assume the initial phase of
  our network, dubbed henceforth $QX_0$, to be generically an
  {\em `almost complete graph'}, i.e. the average number of active
  bonds is assumed to be almost maximal (in other words, $QX_0$ is
  almost a {\em simplex}). That implies that the number of active
  bonds is supposed to be ``near'' $\binom{|V|}{2}$ (with $|V|$ the number of
  nodes). 

Equivalently, we assume the average node degree in the phase $QX_0$ to
remain near its maximum, i.e. $|V|-1$.
\end{assumption}
\begin{defi}[Notations and Conventions] Averages can be taken over the
  network (or graph) at fixed clock time or with respect to clock time
  at e.g. a fixed node or both ``space'' and ``time''. We abbreviate
  the corresponding averaging by
\begin{equation}\langle\cdot\rangle_s\;,\;\langle\cdot\rangle_t\;,\;\langle\cdot\rangle_{st}\end{equation}
\end{defi}
Warning: This kind of averaging should not be confused with the
expectation values defined within the random graph approach (while,
according to our working philosophy, there should be various
interrelations).\\[0.5cm]
Note that, as in cellular automata or, say, statistical
non-equilibrium thermodynamics, the process of averaging has to be
understood in a somewhat pragmatical sense as there does not exist, at
least at the moment, such a stringent and general statistical
framework as in, say, Gibbsian statistical mechanics. In defining
averaged field variables one has to resort to ``sufficiently local''
averages over, on the other side, ``sufficiently many'' microscopic
degrees of freedom. This applies in particular to averaging over clock
time intervals which have to be \tit{``macroscopically small''} but
sufficiently large as compared to the fundamental clock time interval,
$\tau$.
\begin{assumption}[Statistical Version] With the help of the above
  definitions we can formulate our assumption about the initial phase
  as follows: $QX_0$ is assumed to be characterized by
\begin{equation}\langle v(n)\rangle_s\;\text{or}\;\langle
  v(n)\rangle_{st}\;\text{``near''}\;|V|-1\end{equation}
or stated differently
\begin{equation}\langle
  |E|\rangle_t\;\text{``near''}\;\binom{|V|}{2}\end{equation}
\end{assumption}
Evidently there exist certain relations between these notions.
\begin{ob}With 
\begin{equation}\langle v(n)\rangle_s:=\sum_V v(n)/|V|\;\text{and}\;\sum_V
  v(n)/2=|E|\end{equation}
we have
\begin{equation}\langle v(n)\rangle_s\cdot |V|/2=|E|\end{equation}
and
\begin{equation}\langle v(n)\rangle_{st}\cdot |V|/2=\langle E\rangle_t\end{equation}
\end{ob}
\begin{obdef}On the level of the full network phase space (as compared
  with the underlying graph phase space) there are more quantities of
  physical significance, the statistical averages of which describe
  the details of the respective phases the microstate of the network
  is traversing. A case in point is the behavior of the local node
  states. We define
\begin{equation}\langle s\rangle_{\nu},\;\text{with}\;
  \nu:=s,\;st\quad\text{or}\;\langle s_i\rangle_t\end{equation}
\begin{equation}\Delta s:=\langle(s-\langle
  s\rangle)^2\rangle^{1/2}\quad\Delta s_{ik}:=\langle(s_i-s_k)^2\rangle^{1/2}\end{equation}
\begin{equation}\langle(s_i(t+\tau)-s_i(t))^2\rangle^{1/2}\quad\text{etc.}\end{equation}
with the averaging over space, time or space-time respectively.
\end{obdef}

A particularly important role concerning the dynamics of the wiring
diagram is played by $\Delta s_{ik}$ which, as we expect, behaves
similar to $\langle |s_i-s_k|\rangle$. If this quantity is typically
larger than the upper critical parameter of our model dynamics
introduced in section 2 in a certain region of $QX(t)$, then many of the
respective bonds will become temporarily inactive, i.e. are deleted in
$G(t+\tau)$. We plan to make a more detailed statistical or
probabilistic analysis of our model network elsewhere, as such a full
scale probabilistic analysis makes it necessary to investigate quite a
few intricate and technical aspects as e.g. correlation lengths,
the validity of various probabilistic tools in this context like limit
theorems (in the case where the random variables are not completely
independent) and the like.

For the time being, we will be contented with the following brief
analysis. In our network and in particular in the initial phase,
$QX_0$, the average vertex degree $\langle v(n)\rangle$ is expected to
be extremely high. Assuming that in the chaotic initial phase the bond
variables, $J_{ik}$, being incident with an arbitrary node, $n_i$, are
\tit{statistically independent} to a large degree,
we can infer the same for the node states, say, $s_i,s_k$. 
\begin{ob} Assuming the almost statistical independence of bond
  variables being incident with $n_i$ or $n_k$, we can conclude that
  (e.g. for the spacial average) 
\begin{equation}\Delta s = O(\langle v(n)\rangle^{1/2})= O(|V|^{1/2})\end{equation}
in the initial phase $QX_0$. As for the spacial averaging we have
trivially $\langle s_i\rangle=\langle s_k\rangle$ this yields
\begin{equation}\Delta s_{ik}^2\approx \langle s_i\rangle^2+\langle
  s_k\rangle^2,\end{equation}
that is, $\Delta s_{ik}$ fluctuates roughly at the same order 
\begin{equation}\Delta s_{ik}= O(|V|^{1/2})\end{equation}
in the phase $QX_0$.
\end{ob}
\begin{bem}[Almost Statistical Independence] As our network is, for
  the time being, a strictly deterministic system, an assumption like
  complete statistical independence in the above argument would be a
  little bit crude (see also the remarks made in the R\'esum\'e 3.2)
  as there may always exist certain (possibly short-lived)
  correlations between near by degrees of freedom so that something
  like a strict {\em central limit theorem} (see e.g.  \cite{Brei})
  does not hold. But for the above argument to hold, much weaker
  variants would already be sufficient. We gave a preliminary
  discussion of this interesting point in section 3 of \cite{Planck}.
\end{bem}
\begin{conclusion} The above argument shows that it is plausible to
  expect huge fluctuations  on very short scales in the densely
  connected initial phase $QX_0$. If one takes our suggested model of
  a dynamical network law (with the {\em `hystheresis interval
    $(\lambda_1,\lambda_2)$} appropriately tuned) one may entertain the speculative idea
  that under certain favorable circumstances an {\em `avalanche'} may
  be set off in the course of which active bonds are constantly
  thinned out on an increasingly large scale. With this process fairly
  well developed the network may then enter a new geometric phase
  carrying an emergent superstructure we want to associate with the
  discrete statistical substratum of our continuous space-time.
\end{conclusion}
\begin{koro} We think it is not too far fetched to relate this
  hypothetical process with the primordial (pre) big bang scenario, as
  we will show that our unfolding process is accompanied by an
  corresponding increase of macroscopic distance among the
  lumps of the emerging super structure in our network. Furthermore,
  if this picture proves to be correct, it would imply that in our
  expanding universe the infinitesimal physical lumps, representing
  the physical points, are constantly changing both their internal
  structure and external relationships (in a way which may however be
  very slow in the present state of the universe). This will be the
  central issue of the next section.
\end{koro}
Remark: As our main theme is the description and analysis of this new
geometric phase, which is an ambitious task in its own right, we
refrain at this place from going into further details concerning sufficient
or necessary conditions which may trigger this hypothetical phase
transition. In any case we are prepared to change the details of our
model system if this turns out to be necessary (e.g. introduce
bond-bond interactions, to give an example).
\section{The Emergence of Space-Time}
In this core section of our investigation we are going to describe how
the presumed underlying (discrete) fine structure of our continuous
space-time may look like. We want however to emphasize at this place
that, while most of the details and observations we will present are
rigorously proved, the overall picture is still only
hypothetical. That is, we are able to describe a, as we think, fairly
interesting scenario in quite some detail but are not yet able to show
convincingly that our network actually evolves into exactly such a
phase we are going to expound in the following.

In section 3 of this paper and in (\cite{3}, section 3.2) or section 4
of \cite{Planck} we dealt with various properties of (maximal)
subsimplices, $(m)ss$, (also called (maximal) complete subgraphs or
cliques). In \cite{3} we concentrated on their potential geometric
content, in section 3 of this paper we discussed them in the context
of random graphs. Both aspects will now be amalgamated when we
introduce the concept of \tit{`physical points'} and their
\tit{`(causal) entanglement'}.

To begin with, we describe in broad outline our idea of the underlying
discrete substratum of space-time.
\begin{points}[Physical Points]\hfill
\begin{enumerate}
\item Physical points have a (presumably rich) internal structure,
  i.e. they consist of a (presumably) large number of nodes and
  bonds. In the words of Menger they are lumps.
\item We suppose that, what we are used to decribe as fields at a
  space-time point (in fact, rather {\em distributions} in
  e.g. quantum field theory), are really internal excitations of these
  lumps.
\item In order to have a qualitative measure to tell these various
  physical points apart, that is, to discern what happens within a certain point or
  between different points, we conjecture that the individual physical
  points are particularly densely connected subgraphs of our network
  or graph. This then led to our interest in maximal subsimplices.
\item Typically (i.e. if a certain fraction of bonds has been
  eliminated), some of these lumps overlap with each other in a
  stronger or weaker sense, forming so to say {\em `local groups'}
  while other will cease to overlap. This will then establish, in our
  picture, a kind of {\em `proto-causality'} in our {\em
    `proto-space-time'} and will be the central theme of our analysis.
\end{enumerate}
\end{points}

We will proceed by compiling a couple of simple observations
concerning $(m)ss$.
\begin{ob}\hfill
\begin{enumerate}
\item If the node degree, $v_i$, of $n_i$ is
  smaller than $\infty$ then $n_i$ can lie in at most a finite set of
  different simplices, an upper bound being provided by the number of
  different subsets of bonds emerging from $n_i$, that is $2^{v_i}$.
\item The set of subsimplices is evidently `partially ordered' by
  inclusion
\item Furthermore, if S is a simplex, each of its subsets is again a
  simplex (called a `face')
\item It follows that each of the `chains' of linearly ordered
  simplices (containing a certain fixed node) is finite. The corresponding length can be calculated in a
  similar way as in item 1 by selecting chains of sets of bonds,
  ordered by inclusion. In
  other words each chain has a maximal element. By the same token each
  node lies in at least one (but generically several) mss
\item A $mss$ with $n_i$ being a member can comprise at most $(v_i+1)$
  nodes, in other words, its order is bounded by the minimum of these
  numbers when $n_i$ varies over the mss.
\end{enumerate}
\end{ob}
Proof of item 1: Assume that $S_k,S_l$ are two different simplices containing
$n_i$. By definition $n_i$ is linked with all the other nodes in $S_k$
or $S_l$. As these sets are different by assumption, the corresponding
subsets of bonds emerging from $n_i$ are different. On the other side,
not every subset of such bonds corresponds to a simplex (there
respective endpoints need not form a simplex), which proves the upper
bound stated above.\hfill$\Box$.
\begin{ob}The class of simplices, in particular
  the mss, containing a certain fixed node, $n_i$, can be generated in
  a completely algorithmic way, starting from $n_i$. The first level
  consists of the bonds with $n_i$ an end node, the second level
  comprises the triples of nodes ({\em `triangles'}), $(n_in_kn_l)$, with
  the nodes linked with each other and so forth. Each level set can be
  constructed from the preceding one and the process stops when a mss
  is reached.
\end{ob}
Remark: Note that at each intermediate step, i.e. having
already constructed a certain particular subsimplex, one has in general
several possibilities to proceed. On the other hand, a chain of such
choices may differ at certain places from another one but may lead in
the end to the same final simplex (in other words, being simply a permutation of the nodes of the
former simplex).\vspace{0.5cm}

Denoting the $(m)ss$ under discussion by capital $S$ with
certain indices or labels attached to it, this process can be
pictorially abreviated as follows:
\begin{equation}S(n_0\to\cdots\to n_k)\end{equation}
With $S(n_0\to\cdots\to n_k)$ given, each permutation will yield the
same $mss$, i.e:
\begin{equation}S(n_0\to\cdots\to n_k)=S(n_{\pi(0)}\to\cdots\to n_{\pi
    (k)})\end{equation}
Furthermore each $mss$ can be constructed in this way, starting from one of its
nodes. Evidently this could be done for each node and for all possible
alternatives as to the choice of the next node in the above sequence.
\begin{defi} Let $G_{\nu}$ be a class of subgraphs of $G$.\\
\begin{enumerate}
\item $\cap G_{\nu}$ is the graph with $n\in V_{\cap G_{\nu}}$ if $n\in$
every $V_{G_{\nu}}$,\\
$b_{ik}\in E_{\cap G_{\nu}}$ if $b_{ik}\in$ every $E_{G_{\nu}}$\\
\item $\cup G_{\nu}$ is the graph with $n\in V_{\cup G_{\nu}}$ if $n\in
V_{G_{\nu}}$ for at least one $\nu$\\
$b_{ik}\in E_{\cup G_{\nu}}$ if $b_{ik}\in E_{G_{\nu}}$ for at least
one $\nu$.
\end{enumerate}
\end{defi}
As every node or bond belongs to at least one $mss$ (as can be easily
inferred from the above algorithmic construction), we have
\begin{koro}\begin{equation}\cup S_{\nu}=G\end{equation}
\end{koro}

After these preliminary remarks we now turn to our main task, that is,
the analysis of the web of these $mss$ as the elementary building
blocks of the next higher level of organisation. 

In its surmised transition from the almost maximally connected and
chaotic initial phase to the fully developed phase, $QX/ST$ (i.e. $QX$
plus superstructure $ST$), the
underlying graph passes through several clearly distinguishable
epochs. We begin with the epoch where only a small fraction of bonds
is shut off. Let us e.g. assume that $\alpha$ bonds with  
\begin{equation}1\ll\alpha<n/2 \ll n(n-1)/2\quad\text{for $n$ large}\end{equation}
are temporarily dead with $n$ the order of the graph or network
(i.e. the number of nodes) and the rhs of the above equation the
maximal possible number of bonds. In other words, the network is
supposed to be still near the initial phase. We observe that $\alpha$
arbitrarily selected bonds can at most connect $k\le 2\alpha$
different nodes, hence there still exist at least $(n-k)$ nodes
which are maximally connected, viz. they are spanning a still huge
subsimplex $S'\subset G$. On the other hand there are at most $k\le
2\alpha$ nodes with one or more incident bonds missing in the
corresponding induced subgraph.

$V_G$ can hence be split in the following way:
\begin{equation} V_G=V_{S'}\cup V_N\end{equation}
with $V_N$ the unique set of nodes with some of the $\alpha$ bonds among them
missing, $V_{S'}$ the set of remaining nodes being maximally
connected (by construction); i.e. they form a $ss$.
\begin{equation} |V_N|=k\leq 2\alpha\end{equation}
\begin{defi}$[\cup G_i]$ is the induced subgraph spanned by the nodes
  occurring   in $\cup V_i$. Note that in general $[\cup G_i]\supset
  \cup G_i$, that is, it may rather be called its {\em `closure'}.
\end{defi}
\begin{ob}\hfill
\begin{enumerate}
\item The simplex $S'$ is contained in each of
the occurring $mss$, $S_{\nu}$, i.e:
\begin{equation} S'\subset\cap S_{\nu}\;\text{and it holds a fortiori}\;S'=\cap
S_{\nu}\end{equation}
\item Note that $S'$ itself is never maximal since $[S'\cup n_i]$ is always a larger
simplex with $n_i\in N$ and $[S'\cup n_i]$ being the induced subgraph
spanned by $V_{S'}$ and $n_i$.
\item To each maximal simplex $S_{\nu}\subset G$ belongs a unique
maximal subsimplex $N_{\nu}\subset N$ with
\begin{equation} S_{\nu}=[S'\cup N_{\nu}]\end{equation} 
\item It is important for what follows that $S'$ can be uniquely
  characterized, without actually knowing the $S_{\nu}$, by the
  following two properties: i)$S'$ is a $ss$ so that all bonds
  connecting nodes from $V_{S'}$ with $V-V_{S'}$ are ``on''. ii) $S'$ is
  maximal in this class of $ss$, that is, each node in $V-V_{S'}$ has
  at least one bond missing with respect to the other nodes in
  $V-V_{S'}$. An induced subgraph in $G$, having these properties is
  automatically the uniquely given $S'$!
\end{enumerate}
\end{ob}
\begin{koro} From the maximality of the $N_{\nu}$ follows a
general structure relation for the $\{S_{\nu}\}$ and $\{N_{\nu}\}$:
\begin{equation} \nu\neq\mu\,\to\,S_{\nu}\neq S_{\mu}\,\to\,N_{\nu}\neq
N_{\mu}\end{equation}
and neither
\begin{equation} N_{\nu}\subset N_{\mu}\;\mbox{nor}\;N_{\mu}\subset N_{\nu}\end{equation}
viz. there always exists at least one $n_{\nu}\in V_{N_{\nu}}$
s.t. $n_{\nu}\notin V_{N_{\mu}}$ and vice versa.
\end{koro}
Proof of the above observation:
\begin{enumerate}
\item Starting from an arbitrary node $n\in
G$, it is by definition connected with all the nodes in $S'$,
since if say $n,\,n'$ are not connected they both belong to $N$ (by
definition). I.e., irrespectively how we will proceed in the
construction of some $S_{\nu}$, $S'$ can always be added at any
intermediate step, hence $S'\subset\cap S_{\nu}$. On the other side
assume that $n\in \cap S_{\nu}$. This implies that $n$ is connected
with each node in $\cup S_{\nu}$. We showed above that $\cup
S_{\nu}=G$, hence $n$ is connected with all the other nodes, i.e. it
is not in $N$, that is, $n\in S'$, which proves the statement.
\item  As $n\in N$ is connected with each $n'\in S'$ (by definition of
$N$ and $S'$), the subgraph $[S'\cup n]$ is again a (larger)
simplex.
\item We have $S'\subset S_{\nu}$ for all $\nu$, hence 
\begin{equation} S_{\nu}\neq S_{\mu}\;\mbox{implies}\;N_{\nu}\neq N_{\mu}\end{equation}
with $N_{\nu,\mu}$ the corresponding subgraphs in $N$.\\
With $S_{\nu}$ being a simplex, $N_{\nu}$ is again a subsimplex which
is maximal in $N$. Otherwise $S_{\nu}$ would not be maximal in $G$.\\
On the other side each $S_{\nu}=[S'\cup N_{\nu}]$ is uniquely given by
a maximal $N_{\nu}$ in $N$ as each node in $N$ is connected with all
the nodes in $S'$.\end{enumerate}\hfill$\Box$
\vspace{0.5cm}

We see from the above that as long as $\alpha$, the number of dead
(missing) bonds, is sufficiently small, i.e. $2\alpha<n$, there does
exist an overlap $S'$, among the class of $mss$, $S_{\nu}$. This
overlap will become smaller as $\alpha$ increases with clock time $t$;
by the same token the number of $mss$ will increase for a certain
range of the parameter $\alpha$ while the respective size of the $mss$
will shrink..  The above unique characterization of $S'$ in item 4
makes it possible to attack the problem of the order of $S'$ within
the framework of random graphs in a more quantitative manner. Given a
member $G$ of ${\cal G}(n,p)$, $S'$ is fixed by item 4 of the above
observation. We are interested in the probability of $S'$ having, say,
$r$ nodes.

The strategy is, as usual, to try to express the probability of such a
configuration within $G$ as the product of certain more elementary and
(if possible) independent probabilities. Unfortunately this turns out
to be relatively intricate in the above case and we are, at the
moment, only able to provide certain upper and lower bounds for the
probability under discussion. As this example shows that such
questions may not always have simple and straightforward answers, it
is perhaps worthwhile to dwell a little bit on this point.

The typical difficulties one usually encounters in this context are
the following: The structure of the set of graphs in ${\cal G}(n,p)$
having a prescribed property may be rather complicated, so that it is
difficult to avoid multiple counting of members when trying to
calculate the order of such a set. A frequent reason for this is the
intricate entanglement of the various pieces of a complicated graph, a
case in point being the above description of $S'$ in $G$. In our case
the peculiar entanglement can be seen as follows.

Selecting $r$ arbitrary vertices, the probability that the
corresponding induced subgraph forms a $ss$ is $p^{\binom{r}{2}}$ (see
section 3). If this subgraph is to qualify as $S'$, i.e. $S'=\cap
S_{\nu}$, $\cup S_{\nu}=G$, each of the nodes in $N$ is connected with
every node in $S'$. The probability for this property is
$p^{r\cdot(n-r)}$. The difficult part of the reasoning concerns the
subgraph $N$. We call the probability that $N$ has just the structure
being described above, $pr(N)$.

The following observation is helpful. As $S'$ is unique in $G$, i.e.
occurs only once, the corresponding random variable, $X_{S'}$, that we
hit at such an $S'$ of order $r$, when browsing through the set of
induced $r$-subgraphs, is zero with at most one possible
exception, that is, if $S'$ has just the order $r$. Therefore the
corresponding expectation value of $X_{S'}$ is, by the same token,
also the probability of the property ` $S'_r$'.
\begin{conclusion} The probability that a random graph contains such a
 $S'$ of order r is
\begin{equation}pr(S')=\binom{n}{r}\cdot p^{\binom{r}{2}}\cdot
  p^{r(n-r)}\cdot pr(N)\end{equation}
\end{conclusion}
As far as we can see, it is not easy to disentangle $pr(N)$ into more
elementary \tit{independent} probabilities and master the complex
combinatorics. Therefore we will, at the moment, only give (possibly
crude) upper and lower bounds. 

$N$, having $(n-r)$ nodes, is characterized as follows. Labeling the
nodes from $(1)$ to $(n-r)$, none of them is allowed to have the
maximal possible degree (with respect to $N$), i.e. $(n-r-1)$. The
first step is simple. Starting with, say, node $(1)$, the probability
that at least one bond is missing is the complement of the probability
that all possible bonds are present, i.e. $(1-p^{(n-r-1)})$. The
following steps will however become more and more cumbersome. Take
e.g. node $(2)$. In the above probability is already contained both
the probability that either the bond $b_{12}$ is missing or not. If
$b_{12}$ is not missing then some other bond $b_{1i}$ has to be absent
(by the definition of $N$). These two alternatives influence the
possible choices being made at step two. In the former case the
configuration where all bonds $b_{2j},\;j>2$ are present is
admissible, in the latter case this possibility is forbidden.
Depending of which choice we make at each step the algorithmic
construction bifurcates in a somewhat involved manner. Evidently this
first step yields a crude upper bound on $pr(N)$. Making at each step
$(i)$ the particular choice that there are always missing bonds among
the bonds pointing to nodes $(j)$ with $j>i$ provides a lower bound.
We hence have.
\begin{conclusion}
\begin{equation}(1-p^{n-r-1})\geq
  pr(N)\geq\prod_{j=1}^{n-r-1}(1-p^{n-r-j})=\prod_{j=1}^{n-r-1}(1-p^j)\end{equation}
and for $pr(S'=\emptyset)$:
\begin{equation}(1-p^{n-1})\geq
  pr(S'=\emptyset)\geq\prod_{j=1}^{n-1}(1-p^j)\end{equation}
\end{conclusion}
\begin{ob}The lower bound is interesting! Perhaps surprisingly, the
  occurring product is an important {\em number theoretic function}
  belonging to the field of {\em partitions of natural numbers} (see
  any good textbook about combinatorics as e.g. \cite{Halder} or the
  famous book of Hardy and Wright, \cite{Hardy}, the standard source
  being \cite{Andrews}). Our above random graph approach offers the
  opportunity to (re)derive and prove this number theoretic formula by
  purely probabilistic means, i.e. give it an underpinning which seems
  to be, at first glance, quite foreign. We will come back to this
  interesting point elsewhere.
\end{ob}
It is important to have effective estimates for the regime of
probabilities, $p(n)$, so that $S'$ is empty with a high probability.
According to our philosophy this signals the end of the \tit{embryonic
  epoch}, where all the supposed \tit{protopoints} still overlap (and
hence are capable of direct interaction) and the beginning of the
unfolding of a new phase with, as we hope, a more pronounced
\tit{near- and far-order} among the physical points.

Such an estimate can in fact be provided with the help of the above
inequality. We have
\begin{equation}pr(S'=\emptyset)>\prod_1^{n-1}(1-p^j)>\prod_1^{\infty}(1-p^j)=\sum_{k=0}^{\infty}a_kp^k\end{equation}
for $0<p<1$. The following (highly nontrivial) observation is due to
Euler (cf. the above mentioned literature for more recent proofs):
\begin{satz}
\begin{multline}\prod_1^{\infty}(1-p^j)=\sum_{k=0}^{\infty}(-1)^k\cdot\left(p^{\frac{1}{2}(3k^2+k)}+p^{\frac{1}{2}(3k^2-k)}\right)=1-p-p^2+p^5+p^7\ldots\\
\approx
  1-p-p^2\end{multline}
for $p$ small.
\end{satz}
\begin{conclusion}For p near zero, $S'$ is empty with arbitrarily large
  probability $\lesssim 1$.
\end{conclusion}
Remarks: \begin{enumerate}
\item This shows that there exists in fact a regime of small
$p$-values where the embryonic epoch no longer prevails. This holds
the more so for an $n$-dependent $p$ (which is very natural) and
$p(n)\searrow 0$.
\item Note that there exists a possibly substantial class of bond
  configurations which have, up to now, been excluded in the above
  estimate, the inclusion of which would increase the relevant
  probability further.
\end{enumerate}

One can provide other types of estimates concerning the overlap of
$(m)ss$ which will further elucidate this point from other
perspectives. It is particularly desirable to study the detailed structure of
\tit{entanglement} among the $mss$ in the more developed epoch
following the above embryonic scenario. We wil embark on such a
program below, but before doing this, we want
to briefly address two other aspects of the problem. The first one
concerns the explicit construction of a large graph having exactly the
properties we have been talking about. This is also worthwhile for its
own sake as it is frequently rather difficult to construct large
graphs with cetain prescribed properties.

The construction of the example goes as follows. Take $2k$ nodes,
choose a subset $G_1$ consisting of exactly $k$ nodes
$(n_1.\ldots,n_k)$, make $G_1$ a simplex. With the remaining $k$ nodes
$(n_1',\ldots,n_k')$ we proceed in the same way, i.e. we now have two
subsimplices $G_1,G_1'$. We now choose a one-one-map from
$(n_1,\ldots,n_k)$ to $(n_1',\ldots,n_k')$, say:
\begin{equation}n_i\;\to\;n_i'\end{equation}
We now connect all the $n_i$ with the $n_j'$ except for the $k$ pairs
$(n_i,n_i')$. The graph $G$ so constructed has 
\begin{equation}|E_G|=2k(2k-1)/2-k=2k(2k-2)/2\end{equation}
We see from this that, as in our above network scenario, the number of
missing bonds is a relatively small fraction, hence, the example may
be not so untypical.

We can now make the following sequence of observations:
\begin{ob}\begin{enumerate} 
\item $G_1=S_1$ is already a $mss$ as each
$n_i'\in G_1'$ has one bond missing with respect to $G_1$.
\item One gets new $mss$ by exchanging exactly one $n_i$ with its
partner $n_i'$, pictorially:
\begin{equation}[S_1-n_i+n_i']\end{equation}
yielding $k$ further $mss$.
\item One can proceed by constructing another class of $mss$, now
deleting $(n_i,n_j)$ and adding their respective partners, i.e:
\begin{equation}[S_1-n_i-n_j+n_i'+n_j']\end
{equation}
\item This can be done until we end up with the $mss$
\begin{equation}[S_1-n_1-\cdots-n_k+n_1'+\cdots+n_k']=S_1'\end{equation}
The combinatorics goes as follows:
\begin{equation}|\{mss\}|=\sum_{\nu=0}^k {k \choose
    \nu}=(1+1)^k=2^k\end{equation}
i.e., our $2k-$node-graph (with $k$ bonds missing) contains exactly
$2^k$ $mss$ of order $k$.
\item Evidently $S'=\emptyset$
\end{enumerate}
\end{ob}
\begin{ob}\begin{enumerate}\item We showed above that $S'$ is
    non-empty as long as $\alpha$, the number of missing bonds, is
    smaller than $n/2$, $n$ the order of the graph $G$, since $\alpha$
    bonds can at most connect $2\alpha$ different nodes. On the other
    side, $\alpha=n/2$ implies 
\begin{equation}|E_G|=\binom{n}{2}-n/2=n(n-2)/2\end{equation}
or an average vertex degree
\begin{equation}\langle v(n)\rangle_s=n-2\end{equation}
which is still very large.
\item The example constructed in the preceeding observation has
\begin{equation}n=2k\;,\;\alpha=k\;,\;S'=\emptyset\end{equation}
In other words, its parameters are just the ``critical'' ones regarding item 1.
\item In the light of these observations one may surmise that
  $\alpha=n/2$ is perhaps the threshold for $S'=\emptyset$ in the
  sense that, say, 
\begin{equation}pr(S'=\emptyset)=O(1)\end{equation}
for $\alpha\gtrsim n/2$. 

On the other side, within the framework of random graphs, we have
obtained the rigorous but presumably not optimal estimate
\begin{equation}pr(S'=\emptyset)>\prod_1^{\infty}(1-p^j)\approx
  1-p-p^2\end{equation}
for $p$ small. For a large graph $p=1/2$ implies however
\begin{equation}\alpha\approx \frac{1}{2}\binom{n}{2}\gg
  n/2\end{equation}
That is, there is still a wide gap between these two values, a point
which needs further clarification.
\end{enumerate}
\end{ob}

The other point deals with the \tit{diameter} of a typical random
graph and displays the at first glance perhaps surprising phenomenon
that in a wide range of $p$-values, bounded away from zero, this
diameter is only two!
\begin{defi}[Diameter] The diameter, diam(G), of a graph is defined as
  the greatest distance between two arbitrary nodes of the graph, i.e.
\begin{equation}diam(G)=\max_{i,j}d(n_i,n_j)\end{equation}
\end{defi}
Surprisingly, this diameter behaves rather uniformly over a wide range
of $p$-values in a random graph. In order to derive a quantitative
estimate we will proceed as follows. In a first step we calculate the
probability that an arbitrary pair of nodes, $(n_i,n_j)$, is \tit{not}
directly connected with another arbitrary node, called $x$, by a pair
of bonds. For a fixed pair $n_i,n_j$ and $x$ running through the set
of remaining nodes this probability is (following the same line
of reasoning as above) $(1-p^2)^{(n-2)}$. There are $\binom{n}{2}$
such pairs, hence we have (calling the property under discussion $A$)
\begin{equation}pr(A)=\binom{n}{2}\cdot(1-p^2)^{(n-2)}\end{equation}
This has the following consequence:
\begin{ob}The probability that $diam(G)$ is two is $1-pr(A)$. 
\end{ob}
For $n$ large, more precisely, $n\to\infty$, one can now calculate the
$n$-dependent $p^*(n)$-threshold so that $diam(G)=2$ holds almost
shurely for $p(n)>p^*(n)$ (cf. \cite{Random}). We content ourselves
with a simple but nevertheless impressive result. With $p>0$ fixed and
$n\to\infty$ we have from the above estimate
\begin{equation}pr(diam(G)=2)\to 1\end{equation}
\begin{conclusion}With $p>0$ fixed and $n\to\infty$ G(n,p) has
  diameter two almost shurely.
\end{conclusion}
In other words, if one wants to have graphs with large diameter,
which, on the other side, do not fall apart in several pieces, one has
to finetune $p(n)\to 0$ and/or confine the accessible phase space
(dynamically), i.e. let the dynamics select a certain subset of graphs
in $\mcal{G}(n,p)$. This necessity or possibility has been discussed
at length in the preceeding sections and is a feature, physicists are
accustomed to. The result is even less dramatic as, according to our
philosophy, it is rather the connectivity of lumps which interests us
(that is, whole bunches of nodes) and which may be substantially
lower (see below). Furthermore, such a web of node connections, extending over
possibly large distances and lying below the web of lumps which, on
its side, is expected to display a more pronounced near-/far-order, is
actually wellcome as a (in our view) necessary prerequisite of quantum behavior!\\[0.5cm]
Remark: Note that this result is not really surprising even from the
purely mathematical point of view. A $p$ bounded away from zero for
$n\to\infty$ means a  vertex degree scaling roughly as $p\cdot
n$. This implies that on average each node is directly connected with
a substantial fraction of all the nodes in the graph. The situation
is, on the other hand, completely different for, say, \tit{lattice graphs}
which frequently occur in ordinary physics. Here the node degree is low and
constant and remains so for $n\to\infty$, implying that the diameter
diverges.\vspace{0.5cm}

In contrast to the above epoch we surmise the fully developed
space-time superstructure of, say, the present epoch to be organized
in a different way.
\begin{conj}[Fully Developed Space-Time Picture]\hfill
\begin{enumerate}\item Each {\em physical point} or {\em lump},
  identified in our framework with a particular mss, is surrounded by
  a {\em local group} of other lumps, overlapping with it. This
  represents its immediate (infinitesimal) neighborhood. Each of these
  mss, lying in the local group of, say, $S_0$, is again surrounded by
  such a  (slightly different) corona and soforth.
\item Note that the definition of mss does not exclude the possibility
  that there may exist still a substantial fraction of bonds,
  connecting two {\em non-overlapping} mss! Quite to the contrary, we
  consider this to be of tantamount importance as regards the
  incorporation of {\em quantum interaction} (and its, at least in our
  view, hidden {\em non-locality} or {\em entanglement} over
  appreciable macroscopic distances). On the other side, macroscopic
  distance is assumed to be expressed by the relative position of the
  lumps with respect to each other (see below).
\item The whole picture winds up to a {\em two-story} concept of
  space-time. For one there is its macroscopic or quasiclassical
  surface structure, i.e. ordinary almost classical space-time,
  consisting of the relatively smooth carpet of overlapping
  lumps. This granular texture defines, among other things,
  macroscopic distance, dimension and classical causality, i.e. (inter)action
  propagating through ordinary space-time, thus playing rather the
  role of a passive stage.
  
  For another, below this surface there does exists this intricate web
  of more or less randomly wired nodes with bonds, connecting lumps
  which may even be farther apart on the macroscopic scale. Our
  central conjecture is that this weaker and more stochastic
  communication between the various lumps is responsible for the
  quantum effects, we are observing in our present day effective
  quantum theories.
\end{enumerate}
\end{conj}
 
The following abbreviations are useful.
\begin{defi}For $n_i,n_k$ (not being) connected by a bond we write
\begin{equation}n_i\sim n_k\quad (n_i\not\sim n_k)\end{equation}
\end{defi}
We then have
\begin{ob}\hfill\begin{enumerate}
\item $n_i\not\sim n_k$ implies that they are
lying in different $S_{\nu}$'s.
\item $S_{\nu}\,,\,S_{\mu}$ are disjoint, i.e. $S_{\nu}\cap
S_{\mu}=\emptyset$ iff 
\begin{equation} \forall n_{\nu}\in S_{\nu}\;\exists\; n_{\mu}\in
S_{\mu}\;\text{with}\;n_{\nu} \not\sim n_{\mu}\end{equation}
or vice versa.
\end{enumerate}
\end{ob}
\begin{consequences} This shows that it may well be that
$S_{\nu}\cap S_{\mu}=\emptyset$ while the two $mss$ have still a lot of
`{\em interbonds}', i.e. bonds connecting the one with the other. The
guiding idea is however that $V_{S_{\nu}}$ and $V_{S_{\mu}}$, as
a whole, will typically be weaker entangled with
each other than the nodes within $S_{\nu}$ or $S_{\mu}$ when the unfolding process is fully developed.
\end{consequences}
\begin{bem} Note that the superstructure of physical points,
i.e. $\{S_{\nu}\}$, imposes on $G$ a structure called a `\tit{clique
  graph}' with the $S_\nu$ being the supernodes and two supernodes
being adjacent if the corresponding cliques intersect. The
`\tit{infinitesimal neighborhood}', $\mcal{U}_1(S_0)$, of a point
$S_0$ consists of the cliques having non-void intersection with $S_0$.
\end{bem}
\begin{defi} With respect to the above clique graph we can
speak of an\\
\begin{enumerate}
\item `{\em interior bond}' of a given $S_{\nu}$, i.e:
\begin{equation} b_{ik}\;\text{with}\,n_i,\,n_k\,\in S_{\nu}\end{equation}
\item `{\em exterior bond}' with respect to a given $S_{\nu}$, i.e:
\begin{equation} b_{ik}\;\text{with}\,n_i,\,n_k\,\notin S_{\nu}\end{equation}
\item an `{\em interbond}', i.e:
\begin{equation} b_{ik}\;\mbox{with}\,n_i\in S_{\nu},\,n_k\in
S_{\mu},\,\nu\neq\mu\end{equation}
\item a `{\em common bond}' of $S_{\nu}$,$S_{\mu}$ if $b_{ik}$ is an
interior bond both of $S_{\nu}$ and $S_{\mu}$.
\item a `{\it true interbond}' $b_{ik}$ if for $\nu\neq\mu$:
\begin{equation} n_i\in S_{\nu},\,n_k\in S_{\mu},\,n_k\notin S_{\nu}\end{equation}
\item We then have the relation for given $S_{\nu},\,S_{\mu}$:
\begin{equation}
  \{interbonds\}-\{common\;bonds\}=\{true\;interbonds\}\end{equation}
\end{enumerate}
\end{defi}
Remark: The above relations between these classes of bonds describe the (time
dependent) degree of entanglement among the $S_{\nu}$,
that is, among the physical proto points and, as a consequence, the
physical near- and far-order on the level of $ST$, the macroscopic
causality structure of space-time and the non-local entanglement we
observe in quantum mechanics. 
\begin{ob}We now have two (metric) structures on the network or graph,
  the original one with its neighborhood structure and distance
  function, $d(n_i,n_j)$, and the superstructure given by the clique
  graph and its coarse grained neighborhood structure of physical
  points and coarse grained distance function, $d_{cl}(S_i,S_j)$, which we
  regard as a protoform of our ordinary macroscopic distance. Note
  that there may exist a substantial number of interbonds on the lower
  level between supernodes $S_i,S_j$ with $d_{cl}(S_i,S_j)\gg 1$.
\end{ob} 

In the physics of many degrees of freedom what really matters, or
gives ``distance'' a physical content, is not so much some abstract
notion of distance but the strength of interaction or correlation
between the various constituents. In most cases these characteristic
quantities are in fact closely related with their spatial distance in
the system under discussion and its (\tit{intrinsic}) dimension. The
latter one needs however not necessarily be the embedding dimension of
ambient space (cf. e.g. \cite{1})! Following this strand of ideas we
will now organize our network.

Given two node sets $A,\,B$ or the respective subgraphs we can count
the number of bonds connecting them and regard this as a measure of their
mutual dynamical coupling.\\[0.5cm]
\begin{defi}[Connectivity of Subgraphs] With $A,\,B$ being two node
  sets in a given graph, we denote by$|A\sim B|$ the actual number of bonds
  connecting the nodes of $A$ with the nodes of $B$ and by $|A\sim
  B|_m$ the maximal possible number. Then we call
\begin{equation} 0\leq c_{AB}:=|A\sim B|/|A\sim B|_m\leq 1\end{equation}
the `{\it connectivity}' of the pair $A,\,B$. It represents the
probability that a randomly chosen pair of nodes $n_A\in A,\,n_B\in B$
is connected by a bond.  $|A\sim B|_m$ depends however on their
relative position in $G$.
\end{defi}
\begin{ob}\hfill\begin{enumerate}
\item \begin{equation} A\cap
B=\emptyset \to |A\sim B|_m=|A|\cdot|B|\end{equation}
 ($|A|,\,|B|$ the
respective number of nodes), hence 
\begin{equation} c_{AB}=|A\sim
B|/|A|\cdot|B|\end{equation}
\item \begin{equation} A=B\to |A\sim B|_m=\binom{|A|}{2}\end{equation}
\item 
\begin{multline} 
A\cap B\neq\emptyset\to|A\sim B|_m  =  |(A-B)\sim(B-A)|_m\\
                                    +|(A\Delta B)\sim(A\cap
                                   B)|_m+|(A\cap B)\sim(A\cap
                                   B)|_m
\end{multline}
i.e:
\begin{equation} |A\sim B|_m=|A-B|\cdot|B-A|+|A\cap
  B\cdot(|A-B|+|B-A|)+
 \binom {|A\cap B|}{2}\end{equation}
with $A\Delta B$ being the symmetric difference of $A$ and
$B$.
\end{enumerate}
\end{ob}
Remark: There exist of course other possibilities to quantify the
degree of mutual influence among the various regions of the graph. One
could e.g. admit not only bonds but, say, paths up to a certain length
leading from $A$ to $B$. Furthermore there exist a variety of (not
physically motivated) notions of connectedness in graph theory
(e.g. the theorems of Menger, cf. e.g. \cite{11}). The concept
we developed above is adapted to our particular scenario with the
$mss$ as building blocks but with possibly a lot of surviving
interbonds between disjoint grains $S_{\nu},S_{\mu}$.
\begin{conj}[The Metric/Topological Picture of $QX/ST$ (a first Draft)]
The picture we expect to emerge when the unfolding process is fully
developed (i.e. $\alpha$ comparatively large) is now the following.
\begin{enumerate}
\item It is one of the many remarkable observations made in
\cite{Random} that, perhaps against the usual intuition, random graphs
tend to be surprisingly regular, i.e. the {\em generic} graph with, say,
$n$ nodes and $m$ bonds tends to be almost `{\em translation
  invariant}' in an averaged sense, viz. the node degree varies
typically only within a small range (an effect of the peaked
probability in phase space, similar to, say, related phenomena in
statistical mechanics). We learned above that also the typical size of
$mss$ has such a peaked structure. It is perhaps natural to expect the
same for their mutual entanglement.
\item Taking then a typical grain $S_0$ of $ST$, we expect its
infinitesimal neighborhood $\{S^0_{\nu}\}$ to be densely connected
with $S_0$ in the sense of the above definition. In other words: 
\begin{equation}c_{S_0S_0}=1\quad,\quad c_{S_0S^0_{\nu}}\lesssim
  1\end{equation} 
\item Going on in this process of selecting grains, $S_{\mu}$, in
  decreasing order with respect to $c_{S_0S_{\mu}}$ we can construct
  shell after shell around $S_0$ with weaker and weaker connectivity
  as regards to the central element $S_0$, i.e: 
\begin{equation} 1\geq
  c_{S_0S^0_{\nu}}\geq c_{S_0S^1_{\nu'}}\geq c_{S_0S^2_{\nu''}}\geq\cdots\end{equation}
\item We expect (or hope) this process to be consistent with the
  neighborhood structure given on the supergraph $ST$ which is
  defined by intersection of cliques. That is, we hope that node
  distance on $ST$ corresponds more or less with the decrease in
  connectivity in the above sense.
\end{enumerate}
\end{conj}

So far we have described the overall scenario of the phase, $QX/ST$,
in broad outline and have provided a certain arsenal of necessary
conceptual tools, the most important being in our view the definition
of physical points and their entanglement. What remains to be done is,
on the one side, the development of a more detailed and quantitative
underpinning of the still to some extent hypothetical picture. This
concerns in particular the question which class of dynamical network
laws is actually capable to drive the cellular network towards an
attractor, having a structure as depicted by the phase, $QX/ST$. On
the other hand, our general ansatz seems to be surprisingly rich so
that it seems to be not to far fetched to expect possible regions of
overlap to emerge in the future with respect to other fashionable
approaches of a distinctly discrete flavor in quantum gravity.

\end{document}